\documentclass[11pt,twoside]{article}
\usepackage{asp2014}

\aspSuppressVolSlug
\resetcounters

\begin{document}

\title{Novel EOSC Services for Space Challenges: The NEANIAS First Outcomes}

\author{Eva~Sciacca,$^1$ Mel~Krokos,$^3$ 
Ugo~Becciani,$^1$  Cristobal~Bordiu,$^1$ Filomena~Bufano,$^1$
Alessandro~Costa,$^1$
Carmelo~Pino,$^1$ Simone~Riggi,$^1$ 
Fabio~Vitello,$^1$
Carlos~Brandt,$^2$
Angelo~Rossi,$^2$
Eugenio~Topa,$^4$
Simone~Mantovani,$^5$
Laura~Vettorello,$^5$
Thomas~Cecconello,$^6$ and Giuseppe~Vizzari$^6$
\affil{$^1$Istituto Nazionale di Astrofisica (INAF), IT\\
$^2$ Jacobs University Bremen, DE\\
$^3$ University of Portsmouth, UK\\
$^4$ Aerospace Logistics Technology Engineering Company (ALTEC) Spa, IT\\
$^5$ Meteorological Enviromental Earth Observation (MEEO) Srl, IT\\
$^6$University of Milano-Bicocca, IT\\
\email{eva.sciacca@inaf.it}
}}

\paperauthor{Eva~Sciacca}{eva.sciacca@inaf.it}{0000-0002-5574-2787}{INAF}{Osservatorio Astrofisico di Catania}{Catania}{IT}{95123}{Italy}
\paperauthor{Cristobal~Bordiu}{cristobal.bordiu@inaf.it}{0000-0002-7703-0692}{INAF}{Osservatorio Astrofisico di Catania}{Catania}{IT}{95123}{Italy}
\paperauthor{Filomena~Bufano}{filomena.bufano@inaf.it}{0000-0002-3429-2481}{INAF}{Osservatorio Astrofisico di Catania}{Catania}{IT}{95123}{Italy}
\paperauthor{Carmelo~Pino}{carmelo.pino@inaf.it}{}{INAF}{Osservatorio Astrofisico di Catania}{Catania}{IT}{95123}{Italy}
\paperauthor{Simone~Riggi}{simone.riggi@inaf.it}{0000-0001-6368-8330}{INAF}{Osservatorio Astrofisico di Catania}{Catania}{IT}{95123}{Italy}
\paperauthor{Fabio~Vitello}{fabio.vitello@inaf.it}{0000-0003-2203-3797}{INAF}{Istituto di Radioastronomia}{Bologna}{IT}{40129}{Italy}
\paperauthor{Carlos~Brandt}{c.brandt@jacobs-university.de}{0000-0001-6679-3777}{Jacobs University Bremen}{Osservatorio Astrofisico di Catania}{Bremen}{DE}{28759}{Germany}
\paperauthor{Angelo~Rossi}{an.rossi@jacobs-university.de}{}{Jacobs University Bremen}{Osservatorio Astrofisico di Catania}{Bremen}{DE}{28759}{Germany}
\paperauthor{Mel~Krokos}{mel.krokos@port.ac.uk}{0000-0001-5149-6091}{University of Portsmouth}{School of Creative Technologies}{Portsmouth}{UK}{PO12DJ}{United Kingdom}
\paperauthor{Eugenio~Topa}{eugenio.topa@altecspace.it}{}{ALTEC Spa}{}{Torino}{IT}{10146}{Italy}
\paperauthor{Simone~Mantovani}{mantovani@meeo.it }{0000-0003-3979-3645}{MEEO Srl}{}{Ferrara}{IT}{44123}{Italy}
\paperauthor{Laura~Vettorello}{vettorello@meeo.it}{}{MEEO Srl}{}{Ferrara}{IT}{44123}{Italy}
\paperauthor{Thomas~Cecconello}{thomas.cecconello@unimib.it}{}{University of Milano -- Bicocca}{CSAI Research Center}{Milano}{IT}{20126}{Italy}
\paperauthor{Giuseppe~Vizzari}{giuseppe.vizzari@unimib.it}{0000-0002-7916-6438}{University of Milano -- Bicocca}{CSAI Research Center}{Milano}{IT}{20126}{Italy}




  
\begin{abstract}
The European Open Science Cloud (EOSC) initiative faces the challenge of developing an agile, fit-for-purpose, and sustainable service-oriented platform that can address the evolving needs of scientific communities. The NEANIAS project plays an active role in the materialization of the EOSC ecosystem by actively contributing to the technological, procedural, strategic and business development of EOSC. We present the first outcomes of the NEANIAS activities relating to co-design and delivery of new innovative services for space research for data management and visualization (SPACE-VIS), map making and mosaicing (SPACE-MOS) and pattern and structure detection (SPACE-ML). We include a summary of collected user requirements driving our services and methodology for their delivery, together with service access details and pointers to future works. 
\end{abstract}

\section{Introduction}
The European vision for Open Science is driven by the European Open Science Cloud (EOSC) which aims at providing a federated platform to host services supporting all stages of data life cycles in research communities\footnote{\url{https://www.eosc-portal.eu}}. The H2020 NEANIAS\footnote{\url{https://www.neanias.eu/}} project is realising co-design, delivery, and integration into EOSC of new innovative thematic services, founded on state-of-the-art research assets and practices, to underpin underwater, atmospheric and space research. Services are designed to be flexible and extensible to adapt to neighbouring use cases, fostering reproducibility and re-usability.

The NEANIAS space research services are aimed at supporting management and analysis of large data volumes in astrophysics and planetary sciences through \textbf{visualization} (\textbf{\textit{SPACE-VIS}}), efficiently generating large multidimensional \textbf{maps} and \textbf{mosaics} (\textbf{\textit{SPACE-MOS}}), and, finally, supporting mechanisms for automatic detection of structures within maps through \textbf{machine learning} (\textbf{\textit{SPACE-ML}}). We present the first release of these services, reporting on user requirements guiding software developments (section \ref{sec-ur}), describing functionalities (section \ref{sec-ss}) and delivery following the NEANIAS release procedures (section \ref{sec-sd}). Section \ref{sec:conclu} summarises conclusions and outlines future developments.

\section{Space User Requirements}
\label{sec-ur}

Space communities within NEANIAS consist of astrophysics and planetary scientists but also computer scientists and software engineers working on visualisation, vision and machine learning. User requirements have been collected through user stories, to capture real-world workflows of end users and drive software developments via an agile approach shaping up the functionality of the developed services.

User requirements provided by the astrophysics and planetary communities have been formalized in a list of recommendations requesting functionalities to: a) allow multiwavelength studies and improved data access, b) compare images produced with different technical features, improving source identification, classification and characterization in astronomical maps, c) automate image registration and mosaicing for characterising planetary landing  sites and evaluating and predicting mineral resources and, finally d) improve portability, distribution, scalability and performance as well as reproducibility and integrability among services. For more details please see \citep{neaniasd4_1}.

\section{NEANIAS Space Services}
\label{sec-ss}

NEANIAS services are built on top of TRL6 software solutions (i.e. software which is fully functional in its original domain) evolving to TRL8 (i.e. fully operational in cloud environments). 

The \textbf{\textit{SPACE-VIS}} service provides
an integrated operational solution for astrophysics and planetary data management aided by advanced visualization mechanisms, including visual analytics and virtual reality, and is underpinned by FAIR\footnote{\url{https://www.openaire.eu/how-to-make-your-data-fair}} principles.
We exploit ViaLactea to utilise astrophysical surveys to aid understanding of the star formation process of the Milky Way. ViaLactea Visual Analytics (VLVA) combine different types of visualization to perform analysis by exploring correlations \citep{vitello2018vialactea} managed in the ViaLactea Knowledge Base (VLKB). VLKB \citep{molinaro2016vialactea} includes 2D and 3D (velocity cubes) surveys, numerical model outputs, point-like and diffuse object catalogues and allows for retrieval of all available datasets as well as cutouts on the positional and/or velocity axis. The Astra Data Navigator (ADN) is a virtual reality environment
for visualizing large stellar catalogues. The first prototype is described in \citet{lanza2020sviluppo} and has been customised to access cloud services for interactive data exploration and navigation. 
Finally, the Advanced Geospatial Data Management platform (ADAM) \citep{ADAM20} accesses a large variety of environmental data and is customised in NEANIAS to access planetary data provided by PlanetServer\footnote{\url{http://planetserver.eu/}}.

The \textbf{\textit{SPACE-MOS}} service provides tools for making high quality images from raw data (map making) and for assembling such images into custom mosaics (mosaicing). The service exploits Montage\footnote{\url{http://montage.ipac.caltech.edu/}}
and is integrated with VLKB for merging adjacent datasets.
The service also allows integration with data processing pipelines in ADAM\footnote{\url{https://adamplatform.eu}} which offers tools for planetary data analysis and for producing cartographic products, such as Digital Elevation Models (DEMs) and 3D models from stereo imagery.

The \textbf{\textit{SPACE-ML}} service provides advanced solutions for pattern and structure detection in astronomical surveys as well as in planetary surface composition, topography and morphometry. The service integrates cutting-edge machine learning algorithms to perform automatic classification of compact and extended sky structures or planetary surfaces.
We exploit CAESAR \citep{riggi2016} to extract and parametrise compact and extended sources from astronomical radio interferometric maps. The processing pipeline is a series of distinct stages that can run on multiple cores and processors. NEANIAS has integrated a deep learning mechanism to significantly improve source identification, classification and characterisation of sources in large-scale radio surveys.

\section{Services Delivery}
\label{sec-sd}

The first release of the NEANIAS space thematic services targets the NEANIAS ecosystem, currently realised on cloud infrastructures provided by GARR \footnote{\url{https://cloud.garr.it/}} and MEEO\footnote{\url{http://www.meeo.it/cloud/}}. Thematic services are also included on top of core services enabling streamlined integration into EOSC. They support data analysis through Artificial Intelligence (AI) and deep learning and data exploration and discovery in large data volumes through visualisation and virtual reality. 

\begin{figure}
    \centering
    \includegraphics[width=\textwidth]{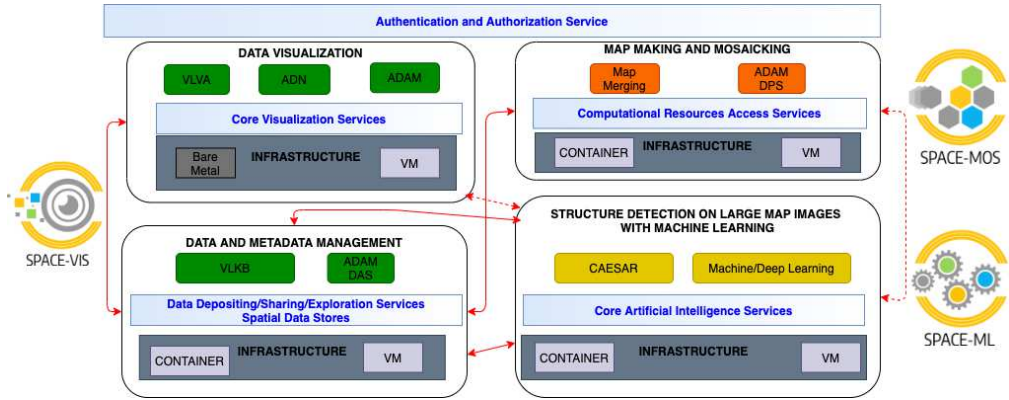}
    \caption{Space Services delivery and integration with core services.}
    \label{fig:spaceservices}
\end{figure}

Figure \ref{fig:spaceservices} outlines the NEANIAS space thematic services delivery and their integration with core services for authentication and authorization, computational and data resources access, as well as visualization and AI services. ViaLactea, ADN and ADAM are released as distributed systems while CAESAR is accessible through REST-APIs on the GARR OpenStack\footnote{\url{https://cloud.garr.it/support/kb/openstack/}} instances and tested on their Kubernetes\footnote{\url{https://cloud.garr.it/support/kb/kubernetes/}} cluster for load scalability connecting to core AI Services.

The NEANIAS space thematic services are reachable through the NEANIAS space portal\footnote{\url{https://thematic.dev.neanias.eu/SPACE/}}. All related documentation has been delivered on the NEANIAS documentation repository\footnote{\url{https://docs.neanias.eu/en/latest/\#space-services}} and the relevant source code is handled by the NEANIAS GitLab repository\footnote{\url{https://gitlab.neanias.eu/}}.

\section{Conclusion and Future Developments}
\label{sec:conclu}

The NEANIAS project is prototyping new innovative space thematic services driving their co-design, delivery, and EOSC integration. The first release of those services has been presented including a summary of core user requirements driving developments and methodology for their delivery. 
Further annual software releases are planned based on comprehensive community surveys \citep{O11-96} and continuous integration mechanisms.

\acknowledgements The research leading to these results has received funding from the European Commissions Horizon 2020 research and innovation programme under the grant agreement No. 863448 (NEANIAS).

\bibliography{P2-16}


\end{document}